\documentclass[preprint2]{aastex}

\def\kms{\ifmmode{\rm km\thinspace s^{-1}}\else km\thinspace s$^{-1}$\fi} 

\def\H{HD 27638}

\def\today{\number\year\space \ifcase\month\or  January\or February\or
        March\or April\or May\or June\or July\or August\or
September\or
        October\or November\or December\fi\space \number\day}

\shortauthors{Torres}
\shorttitle{\H}

\begin{document}

\title{The Multiple System \H}

%\email{************ Draft Version \today\ ************}

\author{Guillermo Torres}

\affil{Harvard-Smithsonian Center for Astrophysics, 60 Garden St.,
Cambridge, MA 02138}
\email{gtorres@cfa.harvard.edu}

\begin{abstract} 

We report spectroscopic observations of \H B, the secondary in a
visual binary in which the physically associated primary (separation
$\sim$19\arcsec) is a \ion{B9}{5} star. The secondary shows strong
Li~$\lambda$6708 absorption suggesting youth, and has attracted
attention in the past as a candidate post-T Tauri star although this
has subsequently been ruled out.  It was previously known to be a
double-lined spectroscopic binary (F8+G6) with a period of 17.6 days,
and to show velocity residuals indicating a more distant massive third
companion with a period of at least 8 years.  Based on our radial
velocity measurements covering more than two cycles of the outer
orbit, along with other measurements, we derive an accurate triple
orbital solution giving an outer period of $9.447 \pm 0.017$ yr. The
third object is more massive than either of the other two components
of \H B, but is not apparent in the spectra. We derive absolute visual
magnitudes and effective temperatures for the three visible stars in
\H.  Isochrone fitting based on those properties gives an age of $200
\pm 50$~Myr for the system. We infer also an inclination angle of
$\sim53\fdg3$ for the inner orbit of \H B. We detect a near-infrared
excess in \H B which we attribute to the third star being a close
binary composed of late-type stars. This explains its large mass and
lack of a visible signature. Modeling of this excess allows us to
infer not only the masses of the components of the unseen companion,
but also the inclination angle of the outer orbit
($\sim$73\arcdeg). The \H\ system is thus at least quintuple.
	
\end{abstract}

\keywords{binaries: spectroscopic --- binaries: visual ---
stars: individual (\H) --- stars: late-type --- techniques: radial
velocities --- techniques: spectroscopic}

\section{Introduction}
\label{sec:introduction}

\H\ (also known as $\chi$ Tau, 59 Tau, HIP 20430, HR 1369, $\alpha =
4^{\rm h} 22^{\rm m} 34\fs94$, $\delta = +25\arcdeg 37\arcmin
45\farcs5$, J2000, $V = 5.395$, SpT = \ion{B9}{5}) is the brighter
component in a visual binary system (ADS 3161, STF 528) with an
angular separation of about 19\arcsec. The relative position of the
companion, \H B ($V = 8.423$, SpT = \ion{G2}{5}), was first recorded
by William Herschel in 1782 \citep[see][]{Lewis:06} and has not
changed much since, indicating the physical association between the
stars. An investigation by \cite{Murphy:69} included this and many
other physical pairs composed of a B-type and a late-type component to
establish the absolute magnitudes of the primaries by reference to the
better known magnitudes of the secondaries. Since the B stars must be
relatively young, the \H\ system was included also in a study by
\cite{Lindroos:86} that concluded that many of the secondaries in such
pairs are post-T Tauri stars, and show other indicators of youth such
as \ion{Ca}{2} H and K emission, H$\alpha$ emission, strong
Li~$\lambda$6708 absorption, strong X-rays, infrared excess, or a
location in the H-R diagram above the Zero Age Main Sequence
(ZAMS). The age of \H A was estimated to be 123 Myr by comparison with
stellar evolution models.

\H B was reported to be a double-lined spectroscopic binary by
\cite{Martin:92}, and independently by \cite{Pallavicini:92}. Both
teams detected significant Li~$\lambda$6708 absorption. The
spectroscopic orbit of the binary with a period of 17.6 days and an
eccentricity of 0.3 was first derived by \cite{Tokovinin:99}, who
pointed out also that the residuals clearly indicated the presence of
a third, rather massive star in the system with a period of a few
years. The elements of the double-lined orbit were published by
\cite{Tokovinin:01}, along with a preliminary solution for the outer
orbit for which $\sim$70\% of the estimated 8-year cycle was
covered. This fit allowed those authors to confirm that the mass of
the third star is larger than either of the objects in the inner pair,
despite there being no sign of it in their spectra.

We report here our own spectroscopic observations of \H B giving full
coverage of the outer orbit over more than two cycles. We model the
properties of the system and find compelling evidence that the third
star is itself a close binary, making the \H\ system at least
quintuple.
	
\section{Spectroscopic observations and reductions}
\label{sec:spectroscopy}

Spectroscopic observations of \H B were conducted at the CfA between
September 1988 and December 2004 (with a 6-year gap from March 1990 to
November 1996) mostly with an echelle spectrograph on the 1.5-m Wyeth
reflector at the Oak Ridge Observatory (Harvard, Massachusetts). A
single order was recorded with an intensified Reticon diode array
giving a spectral coverage of about 45\,\AA\ at a central wavelength
of 5187\,\AA. The resolving power is $\lambda/\Delta\lambda\approx
35,\!000$. Occasional observations were made also with nearly
identical instruments on the 1.5-m Tillinghast reflector at the F.\
L.\ Whipple Observatory (Mt.\ Hopkins, Arizona) and the Multiple
Mirror Telescope (also on Mt.\ Hopkins, Arizona), prior to its
conversion to a monolithic mirror. A total of 72 spectra were
collected, including one archival observation taken at the Tillinghast
reflector much earlier for a different program (October 1983). The
signal-to-noise (S/N) ratios of these observations range from 10 to 40 per
resolution element of 8.5~\kms.

The double-lined nature of the system is evident already in our first
(archival) spectrum from 1983. Radial velocities were derived from all
our observations using TODCOR \citep{Zucker:94}, a two-dimension\-al
cross-correlation algorithm well suited to our relatively low S/N
spectra. TODCOR uses two templates, one for each component of the
binary (hereafter stars Ba and Bb), and significantly reduces
systematics due to line blending that are often unavoidable in
standard one-dimensional cross-correlation techniques \citep[see,
e.g.,][]{Latham:96}.  The templates were selected from a large library
of synthetic spectra based on model atmospheres by R.\ L.\ Kurucz
(available at \url{http://cfaku5.cfa.harvard.edu}), computed for us by
Jon Morse \citep[see also][]{Nordstrom:94,Latham:02}. These calculated
spectra are available for a wide range of effective temperatures
($T_{\rm eff}$), projected rotational velocities ($v \sin i$), surface
gravities ($\log g$) and metallicities. Experience has shown that
radial velocities are largely insensitive to the surface gravity and
metallicity adopted for the templates. Consequently, the optimum
template for each star was determined from grids of cross-correlations
over broad ranges in temperature and rotational velocity
\citep[see][]{Torres:02}, seeking to maximize the average correlation
weighted by the strength of each exposure. Solar metallicity was
assumed to begin with, along with surface gravities of $\log g = 4.5$
for both stars, appropriate for dwarfs. We obtained best fit values
for the temperatures of 6180~K and 5620~K for the primary and
secondary, respectively, with estimated uncertainties of 150~K. These
correspond to spectral types of F8 and G6 \citep{Gray:92}.  The
rotational broadening of the stars was found to be very small
(formally $v \sin i = 1 \pm 3$~\kms\ for both stars), consistent with
the estimates by \cite{Tokovinin:01} of $2.9 \pm 0.6$~\kms\ and
0~\kms\ for the primary and secondary, respectively. We repeated the
procedure for metallicities between [m/H] = $-1.5$ and [m/H] = $+0.5$,
in steps of 0.5 dex, and found the best match to be for solar
composition. This is consistent with the results presented in
\S\ref{sec:system}.

Following \cite{Zucker:94}, we determined also the light ratio between
the secondary and the primary of \H B at the mean wavelength of our
spectroscopic observations (5187\,\AA), which is close to the $V$
band: $\ell_{\rm Bb}/\ell_{\rm Ba} = 0.37 \pm 0.01$. This corresponds
to a magnitude difference $\Delta m = 1.08 \pm 0.03$, and agrees well
with the independent estimate by \cite{Tokovinin:01} of $\Delta m =
1.1$.  Table~\ref{tab:rvcfa} lists the radial velocities for both
components, referred to the heliocentric frame. Typical uncertainties
are given below.  The stability of the zero-point of our velocity
system was monitored by means of exposures of the dusk and dawn sky,
and small systematic run-to-run corrections were applied in the manner
described by \cite{Latham:92}. The accuracy of the CfA velocity
system, which is within about 0.14~\kms\ of the reference frame
defined by minor planets in the solar system, is documented in the
previous citation and also by \cite{Stefanik:99} and \cite{Latham:02}.

\subsection{Orbital solution}
\label{sec:orbit}

The original motivation for our early observations (prior to March
1990) was to investigate systematic errors in the determination of
radial velocities for early-type versus late-type stars by using physical
pairs such as \H AB (D.\ Latham 2005, priv.\ comm.). Although it was
soon realized that no meaningful velocities could be obtained for the
early-type star with our instrumentation, the double-lined nature of the
secondary was certainly noticed and was the reason for collecting
nearly 30 spectra up to that date. However, no orbit was published at
the time. After the implementation of the TODCOR algorithm at the CfA
there was renewed interest in the object as a potentially young
system, and observations were resumed in late 1996.  It then became
obvious that the center of mass of the 17.6-day binary was changing in
response to an additional component, and the system was monitored more
regularly to complete the outer orbit. Only later did we learn of the
independent efforts by \cite{Tokovinin:01} revealing the same trend.

An orbital solution based on all CfA spectra is presented in the
second column of Table~\ref{tab:orbit}. Because the outer period is
much longer than the inner period, we have assumed to first order that
the hierarchical triple system may be separated into an inner orbit
and an outer orbit, the latter being treated as a ``binary'' composed
of the third star (Bc) and the center of mass of the inner pair
(Ba--Bb).  Both orbits were solved simultaneously (12 adjustable
parameters) using standard iterative non-linear least-squares
methods. In particular, we applied the Levenberg-Marquardt technique,
which is described in detail by \cite{Press:92}. The iterations
converged quickly to the final solution, and experiments in which we
varied the initial values of the elements within reason yielded the
same results.  Although the light travel time effect is small, we have
included the appropriate corrections for the inner pair. These depend
on the orbital elements, so they were iterated during the solution.
The outer orbit has a period of 9.4 years, and our data cover 2.2
cycles. The rms residual of the observations, indicative of the
precision of the radial velocities, is 0.59~\kms\ for Ba and
1.03~\kms\ for Bb.

As mentioned earlier, the observations reported originally by
\cite{Tokovinin:01} do not quite cover a full cycle of the outer
orbit. Those measurements were made with a CORAVEL-type
radial-velocity spectrometer \citep[``Radial Velocity Meter",
hereafter RVM;][]{Tokovinin:87} on the 0.7-m telescope at Moscow
University and the 1-m telescope at the Simeis Observatory in Crimea,
and have a precision similar to ours. The observations with this
instrument were continued after publication, and the authors were kind
enough to provide us with an updated list of their RVM velocities for
Ba and Bb. These observations are presented with their permission in
Table~\ref{tab:rvtok}, and now cover nearly one full cycle of the
outer orbit. An independent orbital solution from these data is given
in the third column of Table~\ref{tab:orbit}. There is excellent
agreement with the CfA solution for all of the orbital
elements. Therefore we have combined the measurements, allowing for a
possible velocity offset between the two data sets by incorporating an
additional adjustable parameter ($\Delta RV$). The relative weighting
of the observations was determined by iterations separately for the
primary and secondary components in each data set, based on the
scatter of the measurements.  For the RVM velocities it was found that
the original internal errors (Table~\ref{tab:rvtok}) required the
application of scaling factors of 2.3 and 1.7 to achieve reduced
$\chi^2$ values near unity for stars Ba and Bb, respectively.  The
combined CfA+RVM solution is listed in the final column of
Table~\ref{tab:orbit}. Residuals for the individual CfA and RVM
observations from this fit are given in Table~\ref{tab:rvcfa} and
Table~\ref{tab:rvtok}. The light travel corrections applied to the
dates of the inner pair appear in the last column (the Julian dates
shown are the original values, without corrections). Phases in the
inner and outer orbits are also given.

\begin{figure} 
\vskip -1in
\epsscale{1.0} 
\plotone{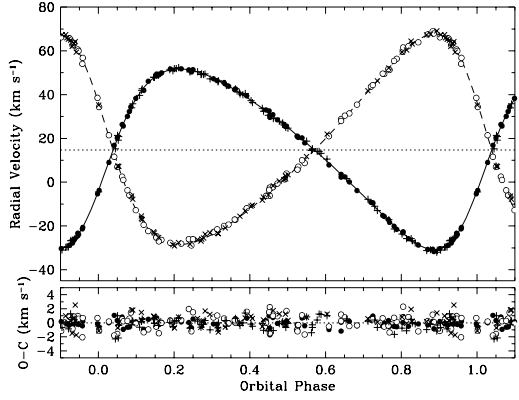}
\vskip -0.2in

 \figcaption[]{Inner orbit of \H B with the motion in the outer orbit
removed. CfA observations are shown with filled circles (primary) and
open circles (secondary), and the RVM measurements are represented
with plus signs (primary) and crosses (secondary). The center-of-mass
velocity is indicated by the dotted line. Residuals are shown at the
bottom. \label{fig:inner}}

 \end{figure}

\begin{figure} 
%\vskip -0.1in
\epsscale{0.95} 
\plotone{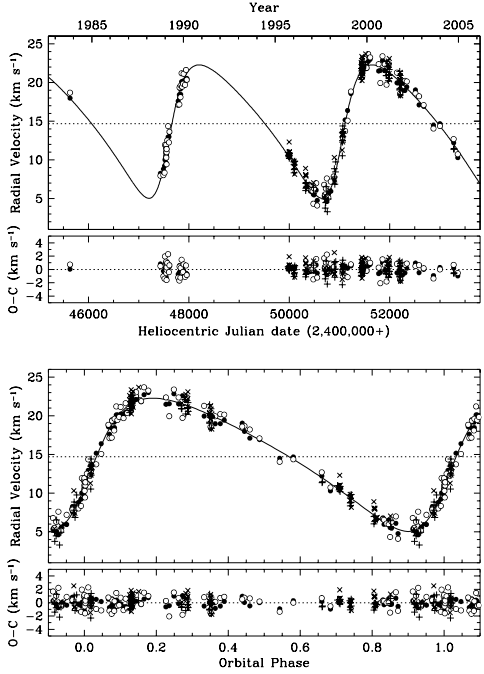}
\vskip 0.5in

 \figcaption[]{Outer orbit of \H B after subtracting the motion in the
 inner orbit from the velocities of each star. Symbols are as in
 Figure~\ref{fig:inner}. The observations and computed orbit are shown
 as a function of time and phase.  Residuals are displayed in the
 smaller panels. \label{fig:outer}}

 \end{figure}

The radial velocities in the inner orbit are presented graphically in
Figure~\ref{fig:inner}, where the motion of the center of mass in the
wide orbit has been subtracted from the individual primary and
secondary velocities. Similarly, the velocities in the outer orbit are
shown in Figure~\ref{fig:outer} both as a function of phase and time,
with the motion in the inner orbit removed. The period of the wide
orbit, $P_{\rm Bab-c} = 9.447 \pm 0.017$ yr, is formally determined
with a relative precision better than 0.2\%.

It is of interest to note that the lines of apsides of the inner and
outer orbits appear to be nearly perfectly aligned, according to our
solution in Table~\ref{tab:orbit}: the difference between the
longitudes of periastron $\omega_{\rm Ba}$ and $\omega_{\rm Bab}$ is
$0\fdg8 \pm 1\fdg5$. This is unlikely to be accidental, and is
presumably a result of dynamical interactions in the triple
system. The outer period happens to be very close to an integer
multiple of the inner period ($P_{\rm Bab-c}/P_{\rm Bab} = 196.03$),
although the formal uncertainty in this ratio (0.32) is large enough
that it is not particularly significant. An improvement in $P_{\rm
Bab-c}$ by an order of magnitude would be needed for a more definitive
conclusion. Of greater interest, as reported by \cite{Tokovinin:01},
is the fact that the mass function (or, equivalently, the minimum mass
of the unseen companion of \H B) is unusually large. For any
reasonable values of the masses of Ba and Bb (see below) the inferred
mass of Bc is larger than either Ba or Bb, yet no light from the third
star is obvious in our spectra.

\cite{Tokovinin:01} made an attempt to detect Bc spectroscopically by
obtaining strong exposures at the phases of maximum velocity
separation between Ba and Bb. No sign of the third star was seen, and
they estimated it must be at least 5 times fainter than Bb. We made a
similar attempt using our CfA spectra, by reprocessing all of them
with an extension of the two-dimensional cross-correlation algorithm
TODCOR to three dimensions \citep{Zucker:95}. Again we failed to
detect star Bc. As suggested by \cite{Tokovinin:01}, one possibility
is that Bc is obscured by circumstellar dust. This would require at
least 2 to 3 magnitudes of extinction in the visible to account for
our inability to see it. Alternatively, Bc could itself be a pair of
late-type stars, or perhaps a massive white dwarf. We examine these
possibilities further below.
	
\section{Properties of the \H\ system}
\label{sec:system}

As mentioned in \S\ref{sec:introduction}, measurements of the relative
position of \H A and \H B over the past 200 years (some 70
observations to date) show that this is a physical pair\footnote{The
total proper motion of the primary as listed in the Tycho-2 Catalog is
$26.1 \pm 1.1$~mas~yr$^{-1}$ \citep{Hog:00}, enough to have changed
the separation by more than 5\arcsec\ in 200 years if \H B were a
background star.}, with only a very slight indication of change
suggesting perhaps orbital motion. Radial velocities for the brighter
star have been measured occasionally but show significant scatter,
possibly due to the large rotational broadening that makes those
measurements difficult \citep[$v \sin i$ measures have ranged between
225~\kms\ and 340~\kms;][]{Abt:02, Uesugi:82}. A representative
average of the radial velocity measures of \H A has been given by
\cite{Barbier:00} as $+15.3 \pm 3.4$~\kms. This is consistent with the
center-of-mass velocity of \H B ($\gamma = +14.694 \pm 0.081$~\kms;
Table~\ref{tab:orbit}), supporting the physical association.  No
trigonometric parallax determination is available for \H B, but \H A
does have an entry in the Hipparcos Catalog \citep{ESA:97} and the
measured parallax is $\pi_{\rm HIP} = 12.19 \pm 1.00$~milli-arc
seconds (mas) corresponding to a distance of about 82~pc.

The brightness of both \H A and \H B has been measured separately in a
variety of photometric systems. The primary is listed in the Hipparcos
Catalog as an `unsolved variable', with a variability amplitude of
$0.036 \pm 0.010$~mag in the $H_p$ passband. No other mention of
variability could be found in the literature. The secondary was
reported as marginally variable by \cite{Lindroos:86} with a maximum
change of $\Delta V = 0.07$~mag based on 4 measurements. However, no
variability appears to have been detected by \cite{Huelamo:04} or
\cite{Martin:92} from more intensive measurements in several
passbands. For our purposes we assume that neither the primary nor the
secondary of the visual pair are significantly variable in brightness.

Adopting the Hipparcos distance of \H A for the system, along with
Johnson $V$-band photometry as compiled by
\cite{Mermilliod:97}\footnote{The values for components A and B are $V
= 5.39$ and $V = 8.42$, based on 9 and 29 individual measurements,
respectively. We assign an uncertainty of 0.01 mag to these values.}
and our light ratio $\ell_{\rm Bb}/\ell_{\rm Ba}$ in
\S\ref{sec:spectroscopy}, we computed absolute visual magnitudes for
the three visible stars: $M_V^{\rm A} = 0.82 \pm 0.18$, $M_V^{\rm Ba}
= 4.20 \pm 0.18$, and $M_V^{\rm Bb} = 5.27 \pm 0.18$.  The errors here
are dominated by the uncertainty in the parallax. Temperatures for
stars Ba and Bb were given in \S\ref{sec:spectroscopy} ($T_{\rm
eff}^{\rm Ba} = 6180 \pm 150$~K and $T_{\rm eff}^{\rm Bb} = 5620 \pm
150$~K). For \H A we have used absolute photometry in the Johnson,
Str\"omgren, and Geneva systems \citep{Mermilliod:97} to derive the
effective temperature based on a large number of color/temperature
calibrations \citep{Popper:80, Moon:85, Gray:92, Napiwotzki:93,
Balona:94, Smalley:95, Kunzli:97, Cox:00}. The various estimates are
in good agreement, and yield an average of $T_{\rm eff}^{\rm A} =
10300 \pm 300$~K, where the uncertainty is a conservative estimate to
account for possible systematics. \cite{Gerbaldi:01} obtained results
consistent with ours. Reddening estimates based on Geneva and
Str\"omgren photometry give negligible values using calibrations by
\cite{Crawford:78} and \cite{Kunzli:97}, in agreement with
\cite{Lindroos:86} and \cite{Gerbaldi:01}.

\begin{figure} 
%\vskip -1.5in
\epsscale{1.25} 
{\hskip -0.5in \plotone{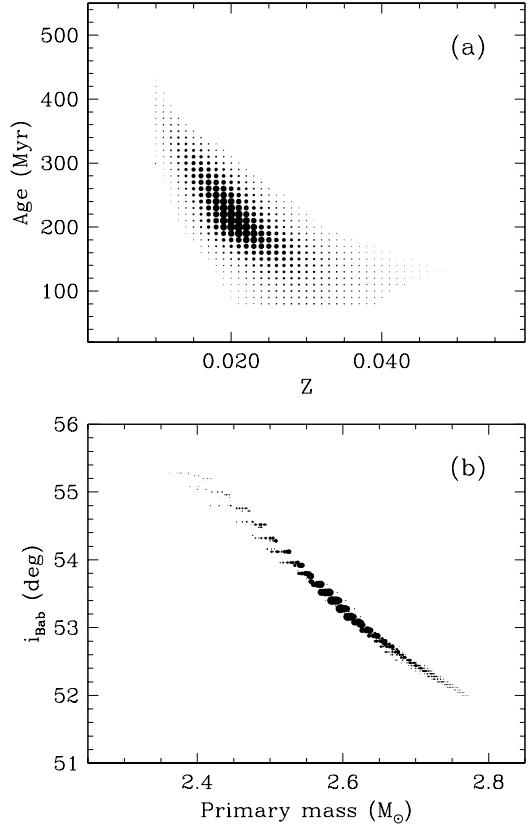}}
\vskip 0.5in

 \figcaption[]{(a) Age and metallicity combinations for the
 Yonsei-Yale isochrones \citep{Yi:01, Demarque:04} that yield a good
 fit the absolute magnitudes and effective temperatures of the three
 visible stars in \H. The size of the points in the shaded region is
 an indication of the quality of the agreement, with larger points
 representing better fits. (b) Mass of the primary ($M_{\rm A}$) and
 inclination angle of the inner orbit in \H B ($i_{\rm Bab}$) for each
 fit in the top panel, showing that these quantities are fairly
 tightly constrained by the best-fitting
 isochrones. \label{fig:metage}}

 \end{figure}

We then used the absolute visual magnitudes and temperatures of the
three visible stars to compare their location in the H-R diagram with
stellar evolution models from the Yonsei-Yale series \citep{Yi:01,
Demarque:04}. We performed a grid search and tested a large number of
isochrones over a range of metal abundances and ages. For each
metallicity/age combination we varied the masses of the stars, seeking
the best match to the observables. With the additional constraint on
the mass ratio for \H B ($q \equiv M_{\rm Bb}/M_{\rm Ba}$), or
equivalently the minimum masses of the components ($M \sin^3 i_{\rm
Bab}$; see Table~\ref{tab:orbit}), we reduced the problem to one of
finding the best values of $M_{\rm A}$ and the inclination of the
inner orbit ($i_{\rm Bab}$) for each age/metallicity combination.  The
result of this grid search is illustrated in Figure~\ref{fig:metage},
where the dots represent models that agree with the observations
within the errors. The best fit isochrone has a metallicity of $Z =
0.021$ and an age of 200~Myr, with uncertainties estimated to be
roughly 0.005 in $Z$ (or 0.1 dex in [Fe/H]) and 50~Myr for the age.
The agreement with the measured magnitudes and temperatures is
excellent (better than 0.1~mag and 20~K for all three stars, which is
well within the errors).  The metallicity is very close to the solar
value in these models ($Z_{\sun} = 0.01812$) and corresponds to [Fe/H]
= $+0.07$.  This is consistent with our crude estimate in
\S\ref{sec:spectroscopy}, and is also not far from the independent
spectroscopic determination for \H B by \cite{Martin:92} of [Fe/H] =
$-0.14$, which, however, may be affected by the double-lined nature of
the object.  The age we infer is somewhat older than previous
estimates, partly due to differences in the models: \cite{Lindroos:86}
estimated 123~Myr, and \cite{Gerbaldi:01} derived
180--190~Myr\footnote{This is their estimate for \H A. The ages they
derived for \H B are much younger (12--19~Myr) because they assumed it
was a post-T Tauri star and used models for the pre-main sequence
phase.}.

The mass of star A is determined to be $M_{\rm A} = 2.60 \pm
0.05$~M$_{\sun}$ and the inclination angle of the Ba--Bb orbit is
$i_{\rm Bab} = 53\fdg3 \pm 0\fdg5$. As a result the individual masses
in \H B are $M_{\rm Ba} = 1.19$~M$_{\sun}$ and $M_{\rm Bb} =
1.02$~M$_{\sun}$. The bottom panel of Figure~\ref{fig:metage} shows
that the values of $M_{\rm A}$ and $i_{\rm Bab}$ are fairly tightly
constrained, as are the age and chemical composition.
Figure~\ref{fig:hr} displays the observations and best-fit isochrone
in the $M_V$/$T_{\rm eff}$ diagram (dashed line).

As a check one may compare the observations with the predictions from
this model for the brightness of the stars in other passbands not used
in the adjustment, in particular in the near-infrared. Observed
magnitudes for \H A and \H B in the $J\!H\!K$ passbands are available from
the 2MASS Catalog, which we have converted to the Johnson system of
\cite{Bessell:88} for consistency with the isochrone computations,
using transformations by \cite{Carpenter:01}.  We find that the
isochrone reproduces the near-infrared colors of \H A fairly well, as
seen in the top third of Table~\ref{tab:irmags}. The differences in
the last column are mostly within the observational errors. On the
other hand, the predictions for \H B (combined light of stars Ba and
Bb, as observed by 2MASS) disagree with the observations in the sense
that the measured colors are redder, and the discrepancy increases
toward longer wavelengths reaching nearly 6$\sigma$ for $V\!-\!K$ (see
middle section of Table~\ref{tab:irmags}). This is strongly suggestive
of a near-infrared excess for \H B, which in all likelihood is related
to the unseen companion.

\section{The nature of the tertiary of \H B}
\label{sec:nature}

The presence of an infrared excess may appear to support the idea
advanced by \cite{Tokovinin:01} that star Bc is perhaps surrounded by
a dust disk that is obscuring the star and preventing its direct
detection.  However, given the age of the system ($\sim$200~Myr), the
possibility that it harbors a circumstellar disk substantial enough to
produce 2--3 magnitudes of optical extinction (see \S\ref{sec:orbit})
is highly unlikely, since such disks in T Tauri stars are known to
dissipate after $\sim$10~Myr \citep[e.g.,][and references
therein]{Mamajek:04}. The age and infrared excess also rule out the
possibility that star Bc is a massive white dwarf. We are left with
the hypothesis that it is instead a binary composed of late-type
dwarfs, and we explore this here by attempting to model the infrared
excess using theoretical isochrones.

\begin{figure} 
%\vskip -1.5in
\epsscale{1.1} 
{\hskip -0.2in \plotone{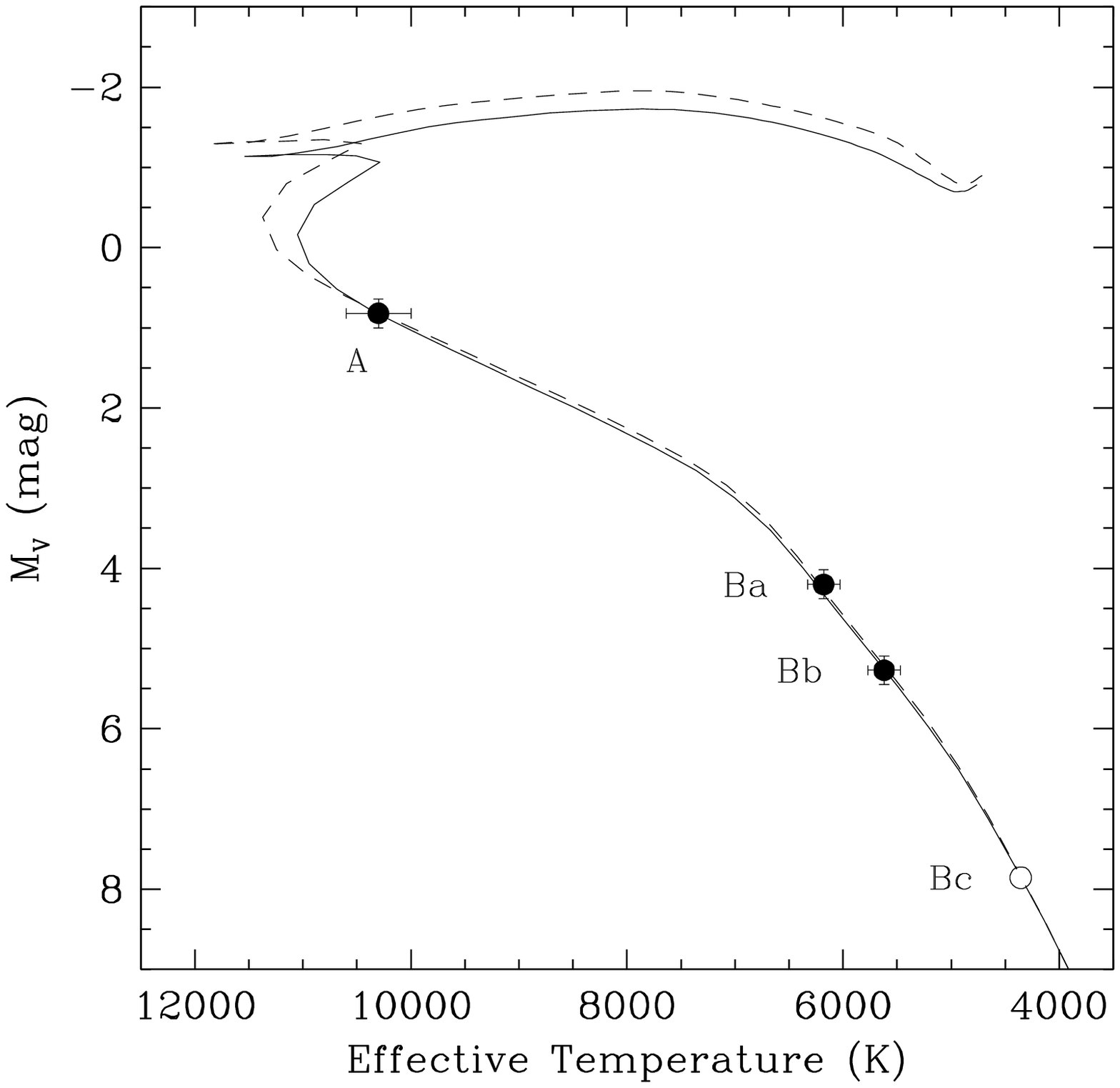}}
\vskip -0.2in

 \figcaption[]{Location of the three visible components of \H\ in the
 H-R diagram (filled circles), along with model isochrones from the
 Yonsei-Yale series \citep{Yi:01, Demarque:04}. The best-fit model ($Z
 = 0.021$, age = 200~Myr) is shown with the dashed line, and a solar
 composition model (age = 240~Myr) used in \S\ref{sec:nature} is
 represented with the solid line and is nearly indistinguishable near
 the observations. The open circle represents the inferred location of
 each component of star Bc (see \S\ref{sec:nature}).\label{fig:hr}}

 \end{figure}

Even though the Yonsei-Yale isochrones used above reach masses as low
as 0.4~M$_{\sun}$, they are not specifically intended for the lower
main-sequence and are thus not expected to reproduce the radiative
properties of real M dwarfs as closely as needed for our
purposes. Other series of models are available that are designed for
low-mass stars and incorporate a more realistic equation of state,
more complete tables of molecular opacities, and non-grey boundary
conditions, all of which have been shown to be critical for this
regime \citep[see, e.g.,][]{Chabrier:97}. In particular, the
\cite{Baraffe:98} models have been found to match the observed colors
of late-type objects fairly well. However, they are not publicly
available for the full range of stellar masses we require, and the
range of metallicities available is also limited. Due to the latter
restriction, we adopt in the following the solar composition for \H,
which is well within the uncertainties of our determination in the
previous section. At this metallicity, the best match to the
magnitudes and temperatures of the three visible stars using the
Yonsei-Yale models is 240~Myr, which is also within the errors and
which we adopt hereafter. This solar-metallicity/240-Myr isochrone is
shown with a solid line in Figure~\ref{fig:hr}, and is nearly
indistinguishable from the $Z = 0.021$/200-Myr model (dashed line)
near the observations.

\begin{figure} 
%\vskip -1.5in
\epsscale{1.1} 
{\hskip -0.1in \plotone{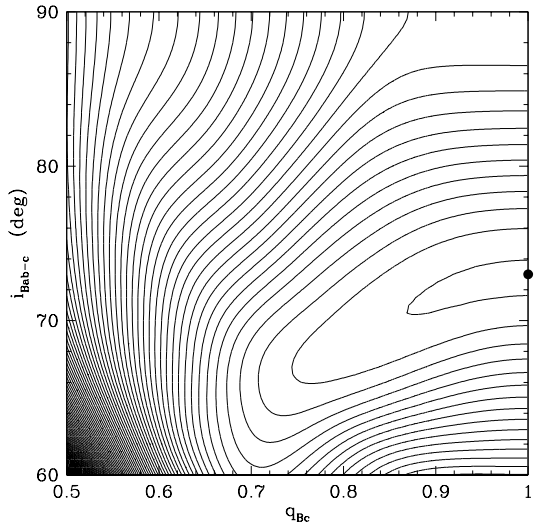}}
\vskip -0.2in

 \figcaption[]{Contour diagram of the $\chi^2$ surface corresponding
 to our modeling of the near-infrared excess of \H B. The variables
 are the mass ratio $q_{\rm Bc}$ of the unseen companion, and the
 inclination angle $i_{\rm Bab-c}$ of the outer orbit. The dot
 indicates the best fit, with $i_{\rm Bab-c} = 73\arcdeg$ and equal
 components for Bc.\label{fig:chi2}}

 \end{figure}

A further complication for our modeling of the infrared excess has to
do with the mixing length parameter, $\alpha_{\rm ML}$. The value
adopted by \cite{Baraffe:98} for solar-type stars is $\alpha_{\rm ML}
= 1.9$, which is the best fit to the properties of the Sun. This value
is presumably also appropriate for stars Ba and Bb in \H, which are
close to a solar mass. However, these models only reach as low as
0.6~M$_{\sun}$ for this $\alpha_{\rm ML}$, and may limit our
predictions for Bc if the components are even smaller. For lower-mass
stars \cite{Baraffe:98} adopt a different mixing length of
$\alpha_{\rm ML} = 1.0$ (with the rest of the physics in the models
being the same), which has been found to give satisfactory results in
modeling the colors of these objects even down to the brown dwarf
regime, but does not reproduce the properties of solar-type stars.
Therefore, we proceeded as follows: We first used an isochrone with
the higher value of $\alpha_{\rm ML}$ to determine the masses of stars
Ba and Bb that provide the best match to their effective temperatures
and absolute visual magnitudes. These masses and the corresponding
infrared colors (which are similar to our previous results using the
Yonsei-Yale models) were then held fixed, and a \cite{Baraffe:98}
isochrone with the same age and metallicity but with $\alpha_{\rm ML}
= 1.0$ was used to explore which combination of masses for the
individual components of Bc gives the best match to the observed
values of $V\!-\!J$, $V\!-\!H$, and $V\!-\!K$ when adding also the
light of Ba and Bb. The observed colors were converted to the CIT
system of \cite{Elias:83} for consistency with the passbands of these
isochrones, following \cite{Carpenter:01}. The masses of the Bc
components are not completely arbitrary: their sum is constrained by
the mass function in the outer orbit (see Table~\ref{tab:orbit}), but
depends also on the inclination angle of that orbit ($i_{\rm
Bab-c}$). Therefore, we performed a grid search varying $i_{\rm
Bab-c}$ and $q_{\rm Bc}$ (the mass ratio between the secondary and the
primary of Bc) over wide ranges, seeking to reproduce the colors. The
best match was found for $i_{\rm Bab-c} = 73\arcdeg \pm 6\arcdeg$ and
equal-mass stars ($q_{\rm Bc} = 1.0$). The corresponding masses for
the Bc components are 0.70~M$_{\sun}$ each. These parameters reproduce
the combined near-infrared colors of \H B very well, as seen in the
lower part of Table~\ref{tab:irmags}. In Figure~\ref{fig:chi2} we show
a contour plot of the $\chi^2$ surface corresponding to the two fitted
variables. The inferred visual brightness of each star in Bc is $M_V =
7.87$, which is about 3.7~mag fainter than star Ba, and explains why
neither we nor \cite{Tokovinin:01} were able to detect them
spectroscopically. The location of Bc in the H-R diagram is shown in
Figure~\ref{fig:hr}.

\section{Discussion and concluding remarks}
\label{sec:discussion}

Our modeling of the infrared excess of \H B shows that the unseen and
over-massive third star is well explained as an equal-mass binary
composed of late-type stars of $M \approx 0.70$~M$_{\sun}$ (spectral
type approximately K4).  As indicated above, the alignment of the
major axes of the inner (Ba--Bb) and outer (Bab--Bc) orbits in this
qquadruple system as evidenced by the nearly identical longitudes of
periastron is a sign that dynamical interactions are at play.  Our
estimates of the inclination angles of the two orbits ($i_{\rm Bab}$
and $i_{\rm Bab-c}$) provide some information on the relative
inclination angle $\phi$, which is of considerable dynamical
importance. That angle is given by $\cos\phi = \cos i_{\rm Bab} \cos
i_{\rm Bab-c} + \sin i_{\rm Bab} \sin i_{\rm Bab-c} \cos(\Omega_{\rm
Bab}-\Omega_{\rm Bab-c})$ \citep[e.g.,][]{Fekel:81}.  The position
angles of the nodes ($\Omega_{\rm Bab}$ and $\Omega_{\rm Bab-c}$) are
unknown, so we can only set limits to $\cos(\Omega_{\rm
Bab}-\Omega_{\rm Bab-c})$ between $-1$ and +1, which leads to $i_{\rm
Bab-c} - i_{\rm Bab} \leq \phi \leq i_{\rm Bab-c} + i_{\rm Bab}$. A
lower limit to $\phi$ can thus be placed at $\phi_{\rm min} =
20\arcdeg \pm 6\arcdeg$, which appears to exclude coplanarity.

With our estimates of $i_{\rm Bab}$ and $i_{\rm Bab-c}$ the total mass
of \H B is 3.6~M$_{\sun}$. The angular semimajor axis of the wide
orbit in this system is therefore $a = 83.6$~mas, corresponding to
6.86~AU. Given the eccentricity and orientation of the orbit, the
angular separation can be as large as 90~mas at times. The combined
brightness of the stars in Bc is expected to be approximately 3.3~mag
fainter than Ba+Bb in $V$, which should make it feasible to resolve
Bc, e.g., by the speckle interferometry technique in the visible on a
4-m class telescope. The brightness difference in the $K$ band is even
more favorable ($\Delta K \approx
1.5$~mag)\footnote{\cite{Tokovinin:01} reported an attempt made in
October 1997 by I.\ I.\ Balega and collaborators to resolve \H B with
the speckle technique in the $K$ band using the 6-m telescope at the
Special Astrophysical Observatory in Zelenchuk, Russia. The companion
was not resolved: the resolution limit was 90~mas, and the predicted
separation according to our orbit was about 60~mas.}.

Perhaps one of the most intriguing properties of \H B is the presence
of strong Li~$\lambda$6708 absorption, as reported originally by
\cite{Gahm:83}, suggesting the star is young. This motivated a number
of studies to explore the possibility that it is a post-T Tauri star.
The equivalent width of the Li line was found to be $156 \pm 4$~m\AA\
by \cite{Martin:92} and 152~m\AA\ by \cite{Pallavicini:92}, although
no account of the double-lined nature of the star was made in these
measurements.  \cite{Pallavicini:92} also reported that no \ion{Ca}{2}
H and K emission is seen in \H B, and that the H$\alpha$ line is in
absorption, which is somewhat unusual for a pre-main sequence
object. Neither of the two visual components have been detected in
X-rays \citep{Schmitt:93, Huelamo:00}. Other than the near-infrared
excess discussed in the present work, which is adequately explained by
binary nature for Bc, no additional excess at longer wavelengths
has been seen in \H B. Although the IRAS satellite detected flux at
12$\mu$m (but not at 25$\mu$m, 60$\mu$m, or 100$\mu$m), that flux is
consistent with being of photospheric origin and does not suggest any
substantial amount of dust \citep{Wyatt:03}. More sensitive
observations by these authors at 850$\mu$m yielded no detection of \H
B. With this evidence and the fact that the age of the system as
estimated from the B-type primary is more typical of ZAMS stars, most
studies have concluded that \H B is not a post-T Tauri star. We
support this conclusion, and our age is in fact even older than
previous estimates. The presence of a strong Li line is a necessary
but not sufficient condition for youth, as pointed out by
\cite{Pallavicini:92}. In fact, this feature is sometimes strong in
evolved RS~CVn binaries and in other old stars \citep[e.g.,][and
others]{Duncan:81, Pallavicini:87, doNascimento:03}, which is
considered to be a result of the interplay between rotation and the
properties of the convective envelopes in these stars, as described in
the previous references. Thus, while typical of very young stars, the
strength of the Li line in \H B is not exceptional for older objects.

From the evidence presented here at least five components are known in
the \H\ system.  The B-type star itself has been examined for
duplicity by the lunar occultation technique \citep{Meyer:95} but was
found to be unresolved. We note, however, that the Hipparcos Catalog
lists the star as a `suspected non-single' \citep{ESA:97} based on the
quality of the astrometric solution, although no convincing non-single
star solution was found. The radial velocity measurements of \H A are
too few in number and too poor in quality to be of much help in this
regard. Thus, it is still possible that additional components are
present, making this a very interesting multiple system.

\acknowledgements 

Thanks are due to J.\ Caruso, R.\ J.\ Davis, D.\ W.\ Latham, R.\ D.\
Mathieu, R.\ P.\ Stefanik, and J.\ Zajac for obtaining most the CfA
spectroscopic observations used in this work, and to R.\ J.\ Davis for
additionally maintaining the CfA echelle database. The author is also
grateful to A.\ A.\ Tokovinin and N.\ A.\ Gorynya for providing their
RVM observations of \H B. We thank the anonymous referee for a prompt
reading of the manuscript and helpful comments.  Partial support for
this work from NSF grant AST-0406183 and NASA's MASSIF SIM Key Project
(BLF57-04) is acknowledged.  This research has made use of the SIMBAD
database, operated at CDS, Strasbourg, France, of NASA's Astrophysics
Data System Abstract Service, of the Washington Double Star Catalog
maintained at the U.S.\ Naval Observatory, and of data products from
the Two Micron All Sky Survey, which is a joint project of the
University of Massachusetts and the Infrared Processing and Analysis
Center/California Institute of Technology, funded by NASA and the NSF.

\clearpage

\begin{deluxetable}{lccccccccc}
\tabletypesize{\scriptsize}
%\tablewidth{11pc}
\tablewidth{0pt}

\tablecaption{CfA radial velocity measurements for \H B in the
heliocentric frame.\label{tab:rvcfa}}

\tablehead{\colhead{HJD} & \colhead{} & \colhead{RV$_{\rm Ba}$} &
           \colhead{$(O-C)_{\rm Ba}$} & \colhead{RV$_{\rm Bb}$} &
           \colhead{$(O-C)_{\rm Bb}$} & \colhead{Inner} &
           \colhead{Outer} & \colhead{$\Delta T$~\tablenotemark{a}} \\
           \colhead{~~~(2,400,000+)~~~} & \colhead{Year} &
           \colhead{($\kms$)} & \colhead{($\kms$)} &
           \colhead{($\kms$)} & \colhead{($\kms$)} & \colhead{Phase} &
           \colhead{Phase} & \colhead{(days)}}

\startdata 
    45631.7736\dotfill &  1983.810 &  $-$19.91 &  $+$0.04 &  $+$62.88 &  $+$0.07 &   0.804 &   0.445  &     $-$0.0172 \\
    47431.7684\dotfill &  1988.739 &  $+$19.13 &  $+$0.83 &  \phn$-$4.72 &  $+$1.43 &   0.063 &   0.967  &     $+$0.0068 \\
    47492.7114\dotfill &  1988.905 &  $+$15.41 &  $-$0.24 &  \phn$+$0.33 &  $-$0.41 &   0.525 &   0.984  &     $+$0.0081 \\
    47514.5937\dotfill &  1988.965 &  $-$23.33 &  $-$0.49 &  $+$46.29 &  $-$0.86 &   0.769 &   0.991  &     $+$0.0085 \\
    47522.8716\dotfill &  1988.988 &  $+$46.62 &  $+$0.04 &  $-$30.74 &  $+$0.07 &   0.239 &   0.993  &     $+$0.0086 \\
    47544.5902\dotfill &  1989.047 &  $+$24.49 &  $-$0.09 &  \phn$-$4.59 &  $-$0.16 &   0.473 &   0.999  &     $+$0.0089 \\
    47545.5598\dotfill &  1989.050 &  $+$16.49 &  $-$0.48 &  \phn$+$3.11 &  $-$0.83 &   0.528 &   1.000  &     $+$0.0089 \\
    47547.5545\dotfill &  1989.056 &  \phn$-$0.31 &  $+$0.07 &  $+$24.96 &  $+$0.12 &   0.641 &   0.000  &     $+$0.0090 \\
    47547.5591\dotfill &  1989.056 &  \phn$-$1.60 &  $-$1.18 &  $+$24.09 &  $-$2.04 &   0.641 &   0.000  &     $+$0.0090 \\
    47549.5592\dotfill &  1989.061 &  $-$19.52 &  $-$0.07 &  $+$45.62 &  $-$0.11 &   0.755 &   0.001  &     $+$0.0090 \\
    47552.7150\dotfill &  1989.070 &  $-$30.49 &  $-$0.16 &  $+$59.53 &  $-$0.28 &   0.934 &   0.002  &     $+$0.0090 \\
    47554.7521\dotfill &  1989.075 &  $+$16.61 &  $+$0.33 &  \phn$+$3.31 &  $+$0.58 &   0.050 &   0.002  &     $+$0.0091 \\
    47555.6199\dotfill &  1989.078 &  $+$34.67 &  $+$0.15 &  $-$16.39 &  $+$0.25 &   0.099 &   0.003  &     $+$0.0091 \\
    47556.5866\dotfill &  1989.080 &  $+$45.83 &  $+$0.70 &  $-$28.30 &  $+$1.21 &   0.154 &   0.003  &     $+$0.0091 \\
    47570.5279\dotfill &  1989.118 &  $-$26.43 &  $+$0.62 &  $+$57.53 &  $+$1.07 &   0.946 &   0.007  &     $+$0.0092 \\
    47585.6424\dotfill &  1989.160 &  $-$25.28 &  $+$0.87 &  $+$58.94 &  $+$1.51 &   0.805 &   0.011  &     $+$0.0094 \\
    47602.5592\dotfill &  1989.206 &  $-$19.14 &  $+$0.32 &  $+$49.71 &  $+$0.56 &   0.766 &   0.016  &     $+$0.0095 \\
    47612.5311\dotfill &  1989.233 &  $+$43.82 &  $+$0.55 &  $-$21.65 &  $+$0.96 &   0.332 &   0.019  &     $+$0.0096 \\
    47787.0247\dotfill &  1989.711 &  $+$53.98 &  $-$0.71 &  $-$25.28 &  $-$1.24 &   0.246 &   0.070  &     $+$0.0089 \\
    47809.9063\dotfill &  1989.774 &  $+$22.11 &  $-$0.42 &  $+$12.88 &  $-$0.72 &   0.545 &   0.076  &     $+$0.0086 \\
    47812.9855\dotfill &  1989.782 &  \phn$-$4.97 &  $+$0.57 &  $+$47.86 &  $+$0.98 &   0.720 &   0.077  &     $+$0.0086 \\
    47832.7845\dotfill &  1989.837 &  $-$24.10 &  $-$0.16 &  $+$68.55 &  $-$0.27 &   0.845 &   0.083  &     $+$0.0083 \\
    47836.6651\dotfill &  1989.847 &  $+$30.38 &  $-$0.93 &  \phn$+$5.74 &  $-$1.62 &   0.066 &   0.084  &     $+$0.0082 \\
    47854.6263\dotfill &  1989.896 &  $+$39.51 &  $+$0.38 &  \phn$-$1.40 &  $+$0.65 &   0.086 &   0.089  &     $+$0.0079 \\
    47864.7828\dotfill &  1989.924 &  \phn$+$5.35 &  $+$0.19 &  $+$36.67 &  $+$0.33 &   0.663 &   0.092  &     $+$0.0078 \\
    47882.7374\dotfill &  1989.973 &  \phn$+$1.70 &  $-$0.41 &  $+$40.95 &  $-$0.71 &   0.683 &   0.097  &     $+$0.0074 \\
    47893.6744\dotfill &  1990.003 &  $+$52.97 &  $-$0.17 &  $-$17.04 &  $-$0.29 &   0.304 &   0.101  &     $+$0.0072 \\
    47923.5507\dotfill &  1990.085 &  \phn$+$2.37 &  $-$0.61 &  $+$40.47 &  $-$1.06 &   0.002 &   0.109  &     $+$0.0066 \\
    47942.5751\dotfill &  1990.137 &  $+$38.60 &  $-$0.59 &  \phn$+$0.47 &  $-$1.02 &   0.082 &   0.115  &     $+$0.0062 \\
    47957.5210\dotfill &  1990.178 &  $-$21.27 &  $-$0.65 &  $+$69.11 &  $-$1.13 &   0.932 &   0.119  &     $+$0.0059 \\
    50408.7920\dotfill &  1996.889 &  $+$43.39 &  $+$0.42 &  $-$36.01 &  $+$0.72 &   0.190 &   0.830  &     $-$0.0075 \\
    50420.7667\dotfill &  1996.922 &  $-$39.14 &  $-$0.31 &  $+$59.12 &  $-$0.54 &   0.870 &   0.833  &     $-$0.0072 \\
    50443.5874\dotfill &  1996.984 &  $+$41.34 &  $+$0.06 &  $-$34.97 &  $+$0.10 &   0.167 &   0.840  &     $-$0.0065 \\
    50464.6277\dotfill &  1997.042 &  $+$33.88 &  $+$0.60 &  $-$25.96 &  $+$1.04 &   0.362 &   0.846  &     $-$0.0059 \\
    50478.6666\dotfill &  1997.081 &  $+$40.10 &  $-$0.26 &  $-$35.95 &  $-$0.44 &   0.160 &   0.850  &     $-$0.0055 \\
    50492.5423\dotfill &  1997.119 &  $-$32.89 &  $-$0.24 &  $+$51.02 &  $-$0.41 &   0.948 &   0.854  &     $-$0.0051 \\
    50535.5573\dotfill &  1997.236 &  $+$30.18 &  $+$0.67 &  $-$21.09 &  $+$1.16 &   0.392 &   0.866  &     $-$0.0038 \\
    50555.5184\dotfill &  1997.291 &  $+$11.41 &  $-$0.55 &  \phn$-$3.65 &  $-$0.96 &   0.526 &   0.872  &     $-$0.0032 \\
    50695.8306\dotfill &  1997.675 &  $+$16.02 &  $+$0.26 &  \phn$-$5.65 &  $+$0.46 &   0.497 &   0.913  &     $+$0.0013 \\
    50711.0249\dotfill &  1997.717 &  $+$32.68 &  $+$0.00 &  $-$26.58 &  $+$0.00 &   0.360 &   0.917  &     $+$0.0018 \\
    50727.8173\dotfill &  1997.763 &  $+$36.98 &  $-$0.21 &  $-$30.97 &  $-$0.37 &   0.314 &   0.922  &     $+$0.0023 \\
    50747.8564\dotfill &  1997.818 &  $+$21.20 &  $-$0.76 &  $-$11.74 &  $-$1.32 &   0.453 &   0.928  &     $+$0.0030 \\
    50777.7853\dotfill &  1997.899 &  $+$39.43 &  $-$0.09 &  $-$34.36 &  $-$0.15 &   0.153 &   0.936  &     $+$0.0039 \\
    50826.5584\dotfill &  1998.033 &  $-$37.15 &  $-$0.38 &  $+$57.43 &  $-$0.66 &   0.924 &   0.951  &     $+$0.0053 \\
    50884.5732\dotfill &  1998.192 &  $+$44.60 &  $+$0.12 &  $-$35.27 &  $+$0.20 &   0.219 &   0.967  &     $+$0.0068 \\
    51062.7874\dotfill &  1998.680 &  $+$42.14 &  $+$0.01 &  $-$21.04 &  $+$0.01 &   0.344 &   0.019  &     $+$0.0096 \\
    51111.8647\dotfill &  1998.814 &  $+$46.10 &  $+$0.50 &  $-$22.33 &  $+$0.87 &   0.132 &   0.033  &     $+$0.0097 \\
    51160.6170\dotfill &  1998.948 &  $-$28.78 &  $+$0.17 &  $+$67.73 &  $+$0.29 &   0.902 &   0.047  &     $+$0.0096 \\
    51226.6126\dotfill &  1999.128 &  \phn$+$5.79 &  $+$0.53 &  $+$33.75 &  $+$0.92 &   0.651 &   0.067  &     $+$0.0090 \\
    51457.8559\dotfill &  1999.761 &  $-$14.58 &  $-$0.56 &  $+$63.34 &  $-$0.98 &   0.788 &   0.134  &     $+$0.0048 \\
    51494.7334\dotfill &  1999.862 &  $-$23.42 &  $+$0.22 &  $+$76.12 &  $+$0.38 &   0.883 &   0.144  &     $+$0.0039 \\
    51540.6589\dotfill &  1999.988 &  $+$33.12 &  $-$0.27 &  \phn$+$8.73 &  $-$0.47 &   0.492 &   0.158  &     $+$0.0028 \\
    51571.6012\dotfill &  2000.073 &  $+$58.94 &  $+$0.54 &  $-$18.52 &  $+$0.94 &   0.250 &   0.166  &     $+$0.0020 \\
    51610.5274\dotfill &  2000.179 &  $+$38.44 &  $+$0.75 &  \phn$+$5.27 &  $+$1.29 &   0.461 &   0.178  &     $+$0.0011 \\
    51779.7801\dotfill &  2000.643 &  $+$37.47 &  $-$0.62 &  \phn$+$3.75 &  $-$1.07 &   0.077 &   0.227  &     $-$0.0032 \\
    51812.8888\dotfill &  2000.733 &  $-$14.00 &  $-$0.45 &  $+$61.41 &  $-$0.78 &   0.958 &   0.236  &     $-$0.0040 \\
    51856.7677\dotfill &  2000.854 &  $+$39.75 &  $+$0.99 &  \phn$+$3.70 &  $+$1.71 &   0.451 &   0.249  &     $-$0.0051 \\
    51901.6080\dotfill &  2000.976 &  \phn$+$1.69 &  $-$0.14 &  $+$45.53 &  $-$0.24 &   0.998 &   0.262  &     $-$0.0061 \\
    51937.5836\dotfill &  2001.075 &  $+$23.71 &  $+$1.06 &  $+$18.43 &  $+$1.84 &   0.042 &   0.273  &     $-$0.0070 \\
    51981.5237\dotfill &  2001.195 &  $+$26.91 &  $+$0.68 &  $+$16.69 &  $+$1.17 &   0.538 &   0.285  &     $-$0.0079 \\
    52144.8816\dotfill &  2001.642 &  $-$20.03 &  $-$0.51 &  $+$67.91 &  $-$0.88 &   0.819 &   0.333  &     $-$0.0113 \\
    52208.8583\dotfill &  2001.818 &  $+$35.00 &  $-$1.71 &  \phn$-$0.80 &  $-$2.96 &   0.453 &   0.351  &     $-$0.0125 \\
    52251.6866\dotfill &  2001.935 &  $-$26.57 &  $-$0.85 &  $+$73.04 &  $-$1.47 &   0.886 &   0.364  &     $-$0.0133 \\
    52302.6350\dotfill &  2002.074 &  $-$15.32 &  $-$0.52 &  $+$60.21 &  $-$0.89 &   0.781 &   0.378  &     $-$0.0141 \\
    52340.5746\dotfill &  2002.178 &  $-$21.75 &  $+$0.19 &  $+$68.17 &  $+$0.32 &   0.936 &   0.389  &     $-$0.0147 \\
    52512.8888\dotfill &  2002.650 &  \phn$-$5.99 &  $+$0.96 &  $+$47.80 &  $+$1.67 &   0.725 &   0.439  &     $-$0.0170 \\
    52583.7527\dotfill &  2002.844 &  $-$12.30 &  $-$0.48 &  $+$52.52 &  $-$0.83 &   0.751 &   0.460  &     $-$0.0177 \\
    52676.6419\dotfill &  2003.098 &  $+$11.24 &  $+$0.01 &  $+$23.71 &  $+$0.01 &   0.028 &   0.487  &     $-$0.0185 \\
    52871.8749\dotfill &  2003.633 &  $+$42.65 &  $-$0.92 &  $-$18.76 &  $-$1.60 &   0.120 &   0.543  &     $-$0.0195 \\
    53000.7831\dotfill &  2003.986 &  $+$32.66 &  $+$0.30 &  \phn$-$6.59 &  $+$0.52 &   0.443 &   0.581  &     $-$0.0196 \\
    53273.8960\dotfill &  2004.733 &  $-$23.33 &  $+$0.20 &  $+$54.13 &  $+$0.35 &   0.959 &   0.660  &     $-$0.0182 \\
    53355.7606\dotfill &  2004.958 &  \phn$+$4.44 &  $-$0.96 &  $+$17.55 &  $-$1.66 &   0.609 &   0.684  &     $-$0.0174 \\
\enddata
\tablenotetext{a}{~Light travel time corrections.}
\end{deluxetable}

\clearpage

\begin{deluxetable}{lccc}
\tabletypesize{\scriptsize}
%\tablenum{2}
\tablecolumns{4}
%\tablewidth{29pc}
\tablewidth{0pc}
\tablecaption{Spectroscopic orbital solutions for \H B.\label{tab:orbit}}
\tablehead{
\colhead{\hfil~~~~~~~~~~~~~~~~~~~~~~~~~~~~~Parameter~~~~~~~~~~~~~~~~~~~~~~~~~~~~~} & \colhead{CfA data} & \colhead{RVM data} & \colhead{Combined} }
\startdata
\sidehead{Adjusted quantities from inner orbit (Ba and Bb)} \\
\noalign{\vskip -6pt}
~~~~$P_{\rm Bab}$ (days)\dotfill                     &  17.602283~$\pm$~0.000039\phn  &  17.60231~$\pm$~0.00016\phn    &  17.602309~$\pm$~0.000036\phn \\
~~~~$K_{\rm Ba}$ (\kms)\dotfill                      &  41.38~$\pm$~0.10\phn          &  41.12~$\pm$~0.13\phn          &  41.287~$\pm$~0.080\phn       \\
~~~~$K_{\rm Bb}$ (\kms)\dotfill                      &  48.15~$\pm$~0.18\phn          &  48.12~$\pm$~0.17\phn          &  48.133~$\pm$~0.080\phn       \\
~~~~$e_{\rm Bab}$\dotfill                            &  0.2950~$\pm$~0.0018           &  0.2929~$\pm$~0.0022           &  0.2938~$\pm$~0.0013          \\
~~~~$\omega_{\rm Ba}$ (deg)\dotfill                  &  249.38~$\pm$~0.44\phn\phn     &  249.60~$\pm$~0.52\phn\phn     &  249.43~$\pm$~0.33\phn\phn    \\
~~~~$T_{\rm Bab}$ (HJD$-$2,400,000)\dotfill          &  48891.644~$\pm$~0.019\phm{2222} &  48891.662~$\pm$~0.028\phm{2222} &  48891.649~$\pm$~0.014\phm{2222}  \\
\sidehead{Adjusted quantities from outer orbit (Ba+Bb and Bc)} \\
\noalign{\vskip -6pt}
~~~~$P_{\rm Bab-c}$ (days)\dotfill                       &  3445.8~$\pm$~8.0\phm{222}     &  3371~$\pm$~58\phn\phn         &  3450.6~$\pm$~6.1\phm{222}    \\
~~~~$K_{\rm Bab}$ (\kms)\dotfill                      &  8.579~$\pm$~0.094             &  8.79~$\pm$~0.11               &  8.625~$\pm$~0.067            \\
~~~~$e_{\rm Bab-c}$\dotfill                              &  0.3459~$\pm$~0.0098           &  0.352~$\pm$~0.016             &  0.3560~$\pm$~0.0068          \\
~~~~$\omega_{\rm Bab}$ (deg)\dotfill                 &  250.8~$\pm$~2.3\phn\phn       &  250.2~$\pm$~2.3\phn\phn       &  250.3~$\pm$~1.5\phn\phn      \\
~~~~$T_{\rm Bab-c}$ (HJD$-$2,400,000)\dotfill            &  47549~$\pm$~14\phm{222}       &  47630~$\pm$~57\phm{222}       &  47546.5~$\pm$~9.7\phm{2222}  \\
\sidehead{Other adjusted quantities} \\
\noalign{\vskip -6pt}
~~~~$\gamma$ (\kms)\dotfill                          &  $+14.673$~$\pm$~0.089\phn\phs &  $+14.502$~$\pm$~0.078\phn\phs &  $+14.694$~$\pm$~0.081\phn\phs\\
~~~~$\Delta RV$(RVM$-$CfA) (\kms)\dotfill             &  \nodata                       &  \nodata                       &  $-0.18$~$\pm$~0.11\phs       \\
\sidehead{Derived quantities from inner orbit} \\						                                                                          
\noalign{\vskip -6pt}								                                                                                          
~~~~$a_{\rm Ba} \sin i_{\rm Bab}$ ($10^6$ km)\dotfill          &  9.569~$\pm$~0.026             &  9.517~$\pm$~0.029             &  9.552~$\pm$~0.017            \\
~~~~$a_{\rm Bb} \sin i_{\rm Bab}$ ($10^6$ km)\dotfill          &  11.137~$\pm$~0.026\phn        &  11.137~$\pm$~0.029\phn        &  11.136~$\pm$~0.017\phn       \\
~~~~$a_{\rm Ba-Bb} \sin i_{\rm Bab}$ (R$_{\sun}$)\dotfill        &  29.750~$\pm$~0.072\phn        &  29.676~$\pm$~0.066\phn        &  29.725~$\pm$~0.045\phn       \\
~~~~$M_{\rm Ba} \sin^3 i_{\rm Bab}$ (M$_{\sun}$)\dotfill       &  0.6140~$\pm$~0.0052           &  0.6110~$\pm$~0.0045           &  0.6130~$\pm$~0.0032          \\
~~~~$M_{\rm Bb} \sin^3 i_{\rm Bab}$ (M$_{\sun}$)\dotfill       &  0.5276~$\pm$~0.0035           &  0.5222~$\pm$~0.0035           &  0.5258~$\pm$~0.0022          \\
~~~~$q \equiv M_{\rm Bb}/M_{\rm Ba}$\dotfill         &  0.8592~$\pm$~0.0041           &  0.8545~$\pm$~0.0041           &  0.8578~$\pm$~0.0026          \\
\sidehead{Derived quantities from outer orbit} \\						                                                                          
\noalign{\vskip -6pt}								                                                                                          
~~~~$a_{\rm Bab} \sin i_{\rm Bab-c}$ ($10^6$ km)\dotfill         &  381.4~$\pm$~5.0\phn\phn       &  381.2~$\pm$~5.3\phn\phn       &  382.5~$\pm$~3.0\phn\phn      \\
~~~~$f(M)$ (M$_{\sun}$)\dotfill                      &  0.1862~$\pm$~0.0072           &  0.1942~$\pm$~0.0069           &  0.1872~$\pm$~0.0043          \\
~~~~$M_{\rm Bc} \sin i_{\rm Bab-c}/(M_{\rm Ba}+M_{\rm Bb}+M_{\rm Bc})^{2/3}$ (M$_{\sun}$)\dotfill &  0.5710~$\pm$~0.0073           &  0.5791~$\pm$~0.0068             &  0.5721~$\pm$~0.0044             \\ 
\sidehead{Other quantities pertaining to the fit} \\						                                          
\noalign{\vskip -6pt}								                                                          
~~~~$N_{\rm obs}$ for Ba\dotfill                     &  72                            &  78                            &  72+78                        \\
~~~~$N_{\rm obs}$ for Bb\dotfill                     &  72                            &  58                            &  72+58                        \\
~~~~Time span (days)\dotfill                         &  7724                          &  3305                          &  7724                         \\
~~~~Cycles in outer orbit\dotfill      &  2.24                          &  0.98                          &  2.24                         \\
~~~~$\sigma_{\rm Ba}$ (\kms)\dotfill                 &  0.59                          &  0.73                          &  0.58 / 0.75                  \\
~~~~$\sigma_{\rm Bb}$ (\kms)\dotfill                 &  1.03                          &  0.93                          &  1.04 / 0.93                  \\
\enddata
\end{deluxetable}

\clearpage
%\LongTables
\begin{deluxetable}{lccccccccccc}
\tabletypesize{\tiny}
%\tablewidth{11pc}
\tablewidth{0pt}

\tablecaption{RVM measurements for \H B in the heliocentric
frame.\label{tab:rvtok}}

\tablehead{\colhead{HJD} & \colhead{} & \colhead{RV$_{\rm Ba}$} & \colhead{$\sigma_{\rm RV}$} &
           \colhead{$(O-C)_{\rm Ba}$} & \colhead{RV$_{\rm Bb}$} & \colhead{$\sigma_{\rm RV}$} &
           \colhead{$(O-C)_{\rm Bb}$} & \colhead{Inner} &
           \colhead{Outer} & \colhead{$\Delta T$~\tablenotemark{a}} \\
           \colhead{~~(2,400,000+)~~} & \colhead{Year} &
           \colhead{($\kms$)} & \colhead{($\kms$)} & \colhead{($\kms$)} &
           \colhead{($\kms$)} & \colhead{($\kms$)} & \colhead{($\kms$)} &\colhead{Phase} &
           \colhead{Phase} & \colhead{(days)}}
\startdata 
    49975.563\dotfill &  1995.703 &  \phn$+$9.83   &   0.24    &   $+$0.40    &  \nodata  &  \nodata  &  \nodata  &    0.578 &   0.704  &  $-$0.0165 \\
    49989.535\dotfill &  1995.741 &  $+$36.62   &   0.32    &   $-$0.05    &  \nodata  &  \nodata  &  \nodata  &    0.372 &   0.708  &  $-$0.0163 \\
    49991.558\dotfill &  1995.747 &  $+$22.62   &   0.27    &   $+$0.16    &  \nodata  &  \nodata  &  \nodata  &    0.487 &   0.709  &  $-$0.0163 \\
    49995.522\dotfill &  1995.758 &  $-$12.30   &   0.30    &   $+$0.24    &  $+$37.34   &  0.65     &  $+$0.59    &    0.712 &   0.710  &  $-$0.0162 \\
    49996.487\dotfill &  1995.760 &  $-$21.65   &   0.37    &   $+$0.16    &  $+$47.66   &  0.57     &  $+$0.13    &    0.767 &   0.710  &  $-$0.0162 \\
    49997.540\dotfill &  1995.763 &  $-$30.28   &   0.30    &   $+$0.51    &  $+$59.92   &  0.52     &  $+$1.93    &    0.826 &   0.710  &  $-$0.0162 \\
    49998.437\dotfill &  1995.766 &  $-$35.35   &   0.34    &   $-$0.17    &  $+$63.16   &  0.48     &  $+$0.07    &    0.877 &   0.711  &  $-$0.0162 \\
    49999.461\dotfill &  1995.769 &  $-$30.61   &   0.24    &   $+$0.49    &  $+$58.99   &  0.32     &  $+$0.68    &    0.936 &   0.711  &  $-$0.0162 \\
    50092.237\dotfill &  1996.023 &  $+$46.53   &   0.33    &   $+$0.28    &  \nodata  &  \nodata  &  \nodata  &    0.206 &   0.738  &  $-$0.0147 \\
    50095.328\dotfill &  1996.031 &  $+$34.48   &   0.52    &   $-$0.03    &  \nodata  &  \nodata  &  \nodata  &    0.382 &   0.739  &  $-$0.0146 \\
    50096.279\dotfill &  1996.034 &  $+$26.99   &   0.30    &   $-$1.08    &  $-$12.30   &  0.54     &  $+$0.42    &    0.436 &   0.739  &  $-$0.0146 \\
    50097.324\dotfill &  1996.036 &  $+$19.56   &   0.24    &   $-$0.67    &  \phn$-$3.96   &  0.47     &  $-$0.35    &    0.495 &   0.739  &  $-$0.0146 \\
    50106.227\dotfill &  1996.061 &  $-$10.00   &   0.44    &   $-$0.49    &  $+$29.71   &  1.09     &  $-$1.16    &    0.001 &   0.742  &  $-$0.0144 \\
    50108.252\dotfill &  1996.066 &  $+$37.10   &   0.34    &   $+$0.53    &  $-$22.60   &  0.51     &  $+$0.29    &    0.116 &   0.742  &  $-$0.0144 \\
    50324.534\dotfill &  1996.659 &  $+$29.58   &   0.31    &   $-$0.18    &  $-$18.65   &  0.55     &  $+$0.92    &    0.403 &   0.805  &  $-$0.0097 \\
    50325.564\dotfill &  1996.661 &  $+$21.34   &   0.36    &   $-$1.11    &  \phn$-$9.33   &  0.75     &  $+$1.74    &    0.462 &   0.805  &  $-$0.0097 \\
    50331.551\dotfill &  1996.678 &  $-$31.49   &   0.47    &   $-$0.77    &  $+$51.60   &  0.75     &  $+$0.81    &    0.802 &   0.807  &  $-$0.0096 \\
    50333.520\dotfill &  1996.683 &  $-$37.49   &   0.54    &   $-$0.02    &  $+$59.60   &  1.07     &  $+$0.98    &    0.914 &   0.808  &  $-$0.0095 \\
    50338.498\dotfill &  1996.697 &  $+$43.10   &   0.40    &   $-$0.55    &  $-$36.38   &  0.72     &  $-$0.33    &    0.197 &   0.809  &  $-$0.0094 \\
    50353.583\dotfill &  1996.738 &  $+$12.81   &   0.31    &   $-$0.31    &  \phn$-$0.78   &  0.56     &  $-$0.01    &    0.054 &   0.814  &  $-$0.0090 \\
    50482.234\dotfill &  1997.090 &  $+$33.59   &   1.15    &   $+$0.65    &  \nodata  &  \nodata  &  \nodata  &    0.362 &   0.851  &  $-$0.0054 \\
    50484.274\dotfill &  1997.096 &  $+$18.75   &   0.33    &   $-$0.10    &  \phn$-$8.59   &  0.67     &  $+$1.28    &    0.478 &   0.851  &  $-$0.0053 \\
    50709.575\dotfill &  1997.713 &  $+$39.48   &   0.29    &   $-$0.21    &  $-$35.25   &  0.47     &  $+$0.26    &    0.278 &   0.917  &  $+$0.0018 \\
    50714.594\dotfill &  1997.726 &  \phn$+$4.74   &   0.23    &   $-$1.40    &  \nodata  &  \nodata  &  \nodata  &    0.563 &   0.918  &  $+$0.0019 \\
    50726.556\dotfill &  1997.759 &  $+$40.78   &   0.22    &   $-$0.77    &  $-$37.91   &  0.57     &  $-$0.41    &    0.242 &   0.922  &  $+$0.0023 \\
    50732.544\dotfill &  1997.776 &  \phn$+$4.49   &   0.23    &   $+$1.25    &  \nodata  &  \nodata  &  \nodata  &    0.583 &   0.923  &  $+$0.0025 \\
    50758.434\dotfill &  1997.846 &  \phn$+$9.77   &   0.47    &   $-$2.18    &  \nodata  &  \nodata  &  \nodata  &    0.053 &   0.931  &  $+$0.0033 \\
    50773.407\dotfill &  1997.887 &  $-$40.29   &   0.49    &   $-$0.66    &  $+$58.13   &  0.86     &  $+$0.17    &    0.904 &   0.935  &  $+$0.0038 \\
    50801.316\dotfill &  1997.964 &  $+$17.29   &   0.62    &   $-$0.12    &  \phn$-$7.73   &  1.07     &  $+$0.07    &    0.490 &   0.943  &  $+$0.0046 \\
    50897.239\dotfill &  1998.227 &  $-$32.68   &   0.33    &   $+$0.01    &  $+$57.13   &  1.01     &  $+$2.53    &    0.939 &   0.971  &  $+$0.0071 \\
    50899.208\dotfill &  1998.232 &  $+$14.32   &   0.51    &   $+$1.01    &  \phn$+$0.36   &  1.06     &  $-$0.72    &    0.051 &   0.972  &  $+$0.0072 \\
    50906.211\dotfill &  1998.251 &  $+$23.71   &   0.55    &   $-$1.17    &  $-$12.65   &  1.00     &  $-$0.62    &    0.449 &   0.974  &  $+$0.0073 \\
    50919.245\dotfill &  1998.287 &  $+$44.85   &   0.97    &   $+$0.07    &  \nodata  &  \nodata  &  \nodata  &    0.189 &   0.977  &  $+$0.0076 \\
    50926.257\dotfill &  1998.306 &  \phn$+$6.91   &   0.59    &   $+$1.19    &  \nodata  &  \nodata  &  \nodata  &    0.588 &   0.979  &  $+$0.0078 \\
    51055.521\dotfill &  1998.660 &  $-$28.52   &   0.49    &   $+$0.75    &  $+$59.87   &  1.13     &  $-$1.48    &    0.931 &   0.017  &  $+$0.0095 \\
    51057.518\dotfill &  1998.665 &  $+$13.21   &   0.26    &   $-$2.29    &  \nodata  &  \nodata  &  \nodata  &    0.045 &   0.018  &  $+$0.0095 \\
    51059.525\dotfill &  1998.671 &  $+$46.85   &   0.49    &   $-$0.39    &  $-$27.27   &  0.81     &  $+$0.28    &    0.159 &   0.018  &  $+$0.0096 \\
    51061.507\dotfill &  1998.676 &  $+$47.23   &   0.32    &   $-$0.64    &  $-$28.77   &  0.61     &  $-$0.63    &    0.271 &   0.019  &  $+$0.0096 \\
    51063.560\dotfill &  1998.682 &  $+$38.66   &   0.37    &   $+$1.38    &  $-$14.94   &  0.78     &  $+$0.70    &    0.388 &   0.019  &  $+$0.0096 \\
    51065.554\dotfill &  1998.687 &  $+$22.85   &   0.27    &   $-$0.07    &  \phn$+$1.60   &  0.53     &  $+$0.35    &    0.501 &   0.020  &  $+$0.0096 \\
    51069.582\dotfill &  1998.698 &  $-$12.64   &   0.31    &   $+$0.40    &  \nodata  &  \nodata  &  \nodata  &    0.730 &   0.021  &  $+$0.0096 \\
    51431.577\dotfill &  1999.689 &  $+$53.72   &   0.34    &   $-$1.00    &  $-$16.59   &  0.61     &  $+$1.19    &    0.295 &   0.126  &  $+$0.0054 \\
    51433.553\dotfill &  1999.695 &  $+$42.94   &   0.25    &   $-$0.50    &  \phn$-$3.96   &  0.65     &  $+$0.63    &    0.407 &   0.126  &  $+$0.0054 \\
    51442.595\dotfill &  1999.720 &  $-$22.97   &   0.40    &   $-$0.87    &  $+$72.00   &  0.57     &  $+$0.02    &    0.921 &   0.129  &  $+$0.0052 \\
    51445.578\dotfill &  1999.728 &  $+$42.46   &   0.21    &   $+$0.27    &  \phn$-$1.08   &  0.44     &  $+$1.83    &    0.091 &   0.130  &  $+$0.0051 \\
    51446.585\dotfill &  1999.731 &  $+$54.56   &   0.26    &   $-$0.07    &  $-$16.51   &  0.42     &  $+$0.89    &    0.148 &   0.130  &  $+$0.0051 \\
    51447.520\dotfill &  1999.733 &  $+$58.57   &   0.37    &   $+$0.27    &  $-$22.38   &  0.62     &  $-$0.73    &    0.201 &   0.131  &  $+$0.0050 \\
    51448.602\dotfill &  1999.736 &  $+$57.24   &   0.27    &   $+$0.24    &  $-$19.52   &  0.50     &  $+$0.61    &    0.262 &   0.131  &  $+$0.0050 \\
    51449.572\dotfill &  1999.739 &  $+$53.05   &   0.23    &   $+$0.00    &  $-$16.81   &  0.40     &  $-$1.32    &    0.317 &   0.131  &  $+$0.0050 \\
    51450.574\dotfill &  1999.741 &  $+$47.00   &   0.28    &   $-$0.38    &  \phn$-$8.84   &  0.47     &  $+$0.03    &    0.374 &   0.131  &  $+$0.0050 \\
    51454.613\dotfill &  1999.753 &  $+$17.28   &   0.31    &   $+$1.02    &  $+$28.70   &  0.66     &  $+$1.23    &    0.604 &   0.133  &  $+$0.0049 \\
    51455.531\dotfill &  1999.755 &  \phn$+$8.44   &   0.32    &   $+$0.58    &  \nodata  &  \nodata  &  \nodata  &    0.656 &   0.133  &  $+$0.0049 \\
    51458.573\dotfill &  1999.763 &  $-$19.96   &   0.35    &   $+$0.00    &  $+$69.97   &  0.58     &  $+$0.20    &    0.829 &   0.134  &  $+$0.0048 \\
    51463.586\dotfill &  1999.777 &  $+$48.82   &   0.36    &   $+$0.19    &  \phn$-$9.69   &  1.06     &  $+$0.43    &    0.114 &   0.135  &  $+$0.0047 \\
    51463.586\dotfill &  1999.777 &  \nodata  &   \nodata &   \nodata  &  \phn$-$8.64   &  0.88     &  $+$1.48    &    0.114 &   0.135  &  $+$0.0047 \\
    51467.487\dotfill &  1999.788 &  $+$51.88   &   0.34    &   $+$0.33    &  $-$14.06   &  0.47     &  $-$0.60    &    0.335 &   0.136  &  $+$0.0046 \\
    51519.356\dotfill &  1999.930 &  $+$56.60   &   0.57    &   $+$0.34    &  $-$16.63   &  1.66     &  $+$1.68    &    0.282 &   0.151  &  $+$0.0033 \\
    51929.192\dotfill &  2001.052 &  $+$22.02   &   0.34    &   $-$0.27    &  \nodata  &  \nodata  &  \nodata  &    0.565 &   0.270  &  $-$0.0068 \\
    51929.192\dotfill &  2001.052 &  $+$21.63   &   0.36    &   $-$0.66    &  \nodata  &  \nodata  &  \nodata  &    0.565 &   0.270  &  $-$0.0068 \\
    51932.210\dotfill &  2001.060 &  \phn$-$6.32   &   0.80    &   $-$0.66    &  $+$53.00   &  0.86     &  $+$0.09    &    0.737 &   0.271  &  $-$0.0068 \\
    51958.179\dotfill &  2001.131 &  $+$58.96   &   0.50    &   $+$0.67    &  $-$21.06   &  0.65     &  $+$0.83    &    0.212 &   0.279  &  $-$0.0074 \\
    51970.185\dotfill &  2001.164 &  $-$25.44   &   0.47    &   $-$1.16    &  $+$75.64   &  0.92     &  $+$1.40    &    0.894 &   0.282  &  $-$0.0077 \\
    51988.187\dotfill &  2001.213 &  $-$23.20   &   0.50    &   $-$0.25    &  $+$72.37   &  0.79     &  $-$0.13    &    0.917 &   0.287  &  $-$0.0081 \\
    51991.196\dotfill &  2001.222 &  $+$40.41   &   0.44    &   $-$0.30    &  \phn$+$0.09   &  0.92     &  $+$1.83    &    0.088 &   0.288  &  $-$0.0082 \\
    51992.220\dotfill &  2001.224 &  $+$54.40   &   0.44    &   $+$0.42    &  $-$16.28   &  1.01     &  $+$0.95    &    0.146 &   0.288  &  $-$0.0082 \\
    51995.210\dotfill &  2001.233 &  $+$52.25   &   0.67    &   $-$0.72    &  $-$16.98   &  0.76     &  $-$0.91    &    0.316 &   0.289  &  $-$0.0082 \\
    52197.602\dotfill &  2001.787 &  $-$20.45   &   0.43    &   $-$1.12    &  $+$65.70   &  0.55     &  $-$0.10    &    0.814 &   0.348  &  $-$0.0123 \\
    52198.544\dotfill &  2001.789 &  $-$25.93   &   0.63    &   $-$0.98    &  $+$72.52   &  0.69     &  $+$0.18    &    0.867 &   0.348  &  $-$0.0123 \\
    52199.594\dotfill &  2001.792 &  $-$21.51   &   0.34    &   $+$1.30    &  $+$70.10   &  0.48     &  $+$0.27    &    0.927 &   0.348  &  $-$0.0124 \\
    52208.507\dotfill &  2001.817 &  $+$38.97   &   0.41    &   $-$0.10    &  \phn$-$2.97   &  0.65     &  $-$0.53    &    0.433 &   0.351  &  $-$0.0125 \\
    52209.580\dotfill &  2001.820 &  $+$31.05   &   0.23    &   $-$0.01    &  \phn$+$8.14   &  0.42     &  $+$1.24    &    0.494 &   0.351  &  $-$0.0125 \\
    52210.558\dotfill &  2001.822 &  $+$23.97   &   1.90    &   $+$0.84    &  $+$16.92   &  1.22     &  $+$0.80    &    0.550 &   0.352  &  $-$0.0126 \\
    52214.640\dotfill &  2001.833 &  $-$16.26   &   1.08    &   $-$1.66    &  \nodata  &  \nodata  &  \nodata  &    0.782 &   0.353  &  $-$0.0126 \\
    52215.611\dotfill &  2001.836 &  $-$22.54   &   0.42    &   $-$0.16    &  \nodata  &  \nodata  &  \nodata  &    0.837 &   0.353  &  $-$0.0126 \\
    52216.637\dotfill &  2001.839 &  $-$25.94   &   1.20    &   $-$0.33    &  \nodata  &  \nodata  &  \nodata  &    0.895 &   0.353  &  $-$0.0127 \\
    52217.469\dotfill &  2001.841 &  $-$20.02   &   0.38    &   $-$0.04    &  $+$64.55   &  0.49     &  $-$1.75    &    0.942 &   0.354  &  $-$0.0127 \\
    52223.285\dotfill &  2001.857 &  $+$54.67   &   0.32    &   $-$0.23    &  $-$20.86   &  0.62     &  $+$0.23    &    0.273 &   0.355  &  $-$0.0128 \\
    53276.539\dotfill &  2004.741 &  $+$36.82   &   0.32    &   $-$0.62    &  \nodata  &  \nodata  &  \nodata  &    0.109 &   0.661  &  $-$0.0182 \\
    53280.565\dotfill &  2004.752 &  $+$41.20   &   0.34    &   $-$0.49    &  \nodata  &  \nodata  &  \nodata  &    0.338 &   0.662  &  $-$0.0182 \\
\enddata
\tablenotetext{a}{~Light travel time corrections.}
\end{deluxetable}

\clearpage

\begin{deluxetable}{lccc}
%\tabletypesize{\tiny}
\tablewidth{0pt}
\tablecolumns{3}
\tablecaption{Near infrared colors for \H A and \H B compared to
model predictions.\label{tab:irmags}}

\tablehead{\colhead{} & \colhead{Observed} & \colhead{Model} &
\colhead{$O\!-\!C$} \\ \colhead{~~~~~Color~~~~~} & \colhead{(mag)} &
\colhead{(mag)} & \colhead{(mag)}}

\startdata
\noalign{\vskip -10pt}
\sidehead{\H A~~\citep[models by][]{Yi:01}\tablenotemark{a}} \\
\noalign{\vskip -15pt}
$M_V$\dotfill     &  \phn0.82~$\pm$~0.18\phn   &  \phs0.81\phn & $+$0.01\phn \\
$V\!-\!J$\dotfill &  $-$0.045~$\pm$~0.035\phs  &  $-$0.064     & $+$0.019    \\
$V\!-\!H$\dotfill &  $-$0.078~$\pm$~0.024\phs  &  $-$0.066     & $-$0.012    \\
$V\!-\!K$\dotfill &  $-$0.094~$\pm$~0.021\phs  &  $-$0.061     & $-$0.033    \\
\sidehead{\H B~~\citep[models by][]{Yi:01}\tablenotemark{a}} \\
\noalign{\vskip -15pt}
$M_V$\dotfill     & \phn3.85~$\pm$~0.18\phn   &  3.87\phn      & $-$0.02\phn \\
$V\!-\!J$\dotfill &    1.133~$\pm$~0.028      &  1.036         & $+$0.097    \\
$V\!-\!H$\dotfill &    1.464~$\pm$~0.033      &  1.347         & $+$0.117    \\
$V\!-\!K$\dotfill &    1.534~$\pm$~0.023      &  1.399         & $+$0.135    \\
\sidehead{\H B~~\citep[models by][]{Baraffe:98}\tablenotemark{b}} \\
\noalign{\vskip -15pt}
$M_V$\dotfill     & \phn3.85~$\pm$~0.18\phn   &  3.81\phn      & $+$0.04\phn \\
$V\!-\!J$\dotfill &    1.183~$\pm$~0.026      &  1.180         & $+$0.003    \\
$V\!-\!H$\dotfill &    1.483~$\pm$~0.030      &  1.501         & $-$0.018    \\
$V\!-\!K$\dotfill &    1.554~$\pm$~0.023      &  1.542         & $+$0.012    \\
\enddata

\tablenotetext{a}{~Infrared magnitudes are in the Johnson system of
\cite{Bessell:88}. The modeling for \H B does not account for the
unseen companion.}

\tablenotetext{b}{~Infrared magnitudes are in the CIT system of
\cite{Elias:83}. The modeling in this case accounts for the unseen
companion, and assumes it is a binary composed of late-type dwarfs
(see text).}

\end{deluxetable}

\clearpage

%\clearpage

%\clearpage

%\clearpage

%\clearpage

\end{document}